# Digital Grid: Transforming the Electric Power Grid into an Innovation Engine for the United States

Aranya Chakrabortty and Alex Huang
North Carolina State University

The electric power grid is one of the largest and most complex infrastructures ever built by mankind. Modern civilization depends on it for industry production, human mobility, and comfortable living. However, many critical technologies such as the 60 Hz transformers were developed at the beginning of the 20th century and have changed very little since then.[1] The traditional unidirectional power from the generation to the customer through the transmission-distribution grid has also changed nominally, but it no longer meets the need of the 21st century market energy customers. On one hand, 128m US residential customers pay $15B/per month for their utility bill, yet they have no option to select their energy supplier. In a world of where many traditional industries are transformed by digital Internet technology (Amazon, Ebay, Uber, Airbnb), the traditional electric energy market is lagging significantly behind. A move towards a true digital grid is needed. Such a digital grid requires a tight integration of the physical layer (energy and power) with digital and cyber information to allow an open and real time market akin to the world of e-commerce. Another major factor that is pushing for this radical transformation are the rapidly changing patterns in energy resources ownership and load flow. Driven by the decreasing cost in distributed solar, energy storage, electric vehicle, on site generation and microgrids, the high penetration of Distributed Energy Resource (DER) is shifting challenges substantially towards the edge of grid from the control point of view. The envisioned Digital Grid must facilitate the open competition and open innovation needed to accelerate of the adoption of new DER technologies while satisfying challenges in grid stability, data explosion and cyber security.

The American Recovery and Reinvestment Act of 2009 was an unprecedented action to stimulate the economy and to modernize our nation's energy and communication infrastructure and enhance energy independence. Under this act, the US Department of Energy and the electricity industry have jointly invested in 99 cost-shared projects involving more than 200 participating electric utilities and other organizations to modernize the electric grid, improve inter-operability, and collect an unprecedented level of data on smart grid operations and benefits. A new and substantial investment is needed to develop the new Digital Grid. **Radical reductions in the cost of sensors, communication, information processing, cyber-security, and new regulatory policies in electricity markets are needed**. **Federal leadership is essential** to break the logjam and encourage new approaches. The key is designing programs that drive change but leave plenty of room for ideas from unexpected sources.

**Challenges and Potential Solution**
There are many challenges to transform the power grid into a Digital Grid. The primary ones are listed as follows:

**1.      Electricity Market and Regulatory Policy**
Low marginal costs coupled with high initial capital costs are characteristic of traditional power generation. Today the ratio between running cost and capital cost is falling rapidly with the introduction of renewable generation, along with flexibility of resources such as storage and demand response. Standard economic theory predicts that price of power will collapse in this setting, which is indeed the case today. Wholesale prices for electric power in the U.S. and other parts of the world are frequently far below operating cost, and even negative prices are commonly observed. This trend and price volatility will discourage capital investment, which is why FERC and local authorities are searching for market designs that will encourage long-term investment. A holistic approach to market design that accommodates both surplus and shortage in renewables, combines them with other generation assets as well as storage, and includes design guidelines and incentives that enable short-term benefits of improved grid performance and long-term benefits of sustainability and economic stability is highly warranted. Innovations in wholesale markets need to be accompanied by those in retail markets to accommodate multiple DG owners and aggregators.

On the demand side, changes are needed due to the same reason as above, to accommodate intermittency and uncertainty in renewable generation. The standard practice of generation following load has to be shed, and loads need to be more dispatchable as well. This approach leads to the generalized concept of Demand Response, which needs to be integrated into the functioning of electricity markets. Direct Load Control and Transactive Control, the two methods that make up Demand Response are both highly promising methods that can help in shaping demand so as to coordinate fluctuations in renewables. Research is needed to ensure that consumer engagement will enhance rather than disrupt reliability of the grid. Further, automation of demand response is required to ensure grid reliability and also to ensure that each participant receives the benefits that are promised. For example, strict bounds on indoor temperature and humidity must be maintained. Market rules are required to ensure that players receive transparent incentives for services provided to the grid, even in the case of residential consumers who have little understanding of power systems or even their own fuse box. Given that all of the aforementioned innovations require real-time and reliable communication, a **cyber-physical approach is essential** to realize an efficient electricity market.

---

[1] Mynatt et al. (2017) "A National Research Agenda for Intelligent Infrastructure" CCC Led Whitepapers
http://cra.org/ccc/resources/ccc-led-whitepapers/, last accessed April 12, 2017.

2. **Grid Stability and Resiliency under High DER Penetration**

The envisioned Digital Grid must be realized with no comprise to grid reliability and resilience. Since the Digital Grid is fundamentally supporting high DER penetration with very dynamic bidirectional power flows affected by the real-time market participation, the grand challenges in control is to achieve voltage and frequency stability, as well as intelligent protection. To address these issues as well as to the severe intermittence issue in the DER and the lack of traditional system inertia, the Digital Grid will need the development and installation of a new generation of Smart Transformers as an universal cyber-physical system interface, that can provide the needed system inertia and provide autonomous voltage, frequency and protection. It must also have the secured communication and software such as the Block-chain to participate in the real time energy market for resource sharing (a key attribute to the Digital Grid) by aggregating the DER, load and storage devices. New transmission technologies such as the High Voltage Direct Current (HVDC) must be further developed and installed to connect highly dynamic distribution grids with strong regional interconnection to balance of the system demand and supply.

3. **Data Analytics**

Availability of smart grid data is expected to increase precipitously. Compared to the traditional Supervisory Control and Data Acquisition (SCADA) systems available for operations, these future data sources could provide vastly richer information about the state of the grid. Transforming this raw data into contextual information to support operations requires new data platforms and analytics capable of extracting useful information. This new information will enable utilities to: 1) create new control paradigms that operate the grid closer to system limits, lowering the cost of operation without sacrificing reliability; 2) improve asset utilization and management across transmission and distribution systems; and 3) integrate high levels of renewable generation. The unique characteristics of the new information available to operations create a number of challenges, including: extensive geo-spatial distribution, temporal diversity, internal organizational boundaries, disparate underlying physical sources and a wide range of data quality. Despite extensive investments in various IT technologies, utilities still lack sufficient infrastructure and understanding of the best approaches to manage this data. Major challenges for utilities today include data quality, data volume, data velocity, and especially data variety, although they clearly have issues to address in all these dimensions.

4. **New Communication Technologies for Smart Grid**

Currently one of the biggest roadblocks for grid modernization is that the Information Technology (IT) infrastructure for today's grid is rigid and low capacity. The push to adopt the existing open Internet and high-performance computing technologies would not be enough to meet the requirement of collecting and processing very large volumes of real-time data. Secondly, the impact of unreliable and insecure communication and computation infrastructures on grid operations is not well understood. Typically, the Internet cannot provide the required latency and packet loss performance for grid operation under high data-rates. Moreover, the network performance is highly random, and therefore, difficult to model accurately. Few studies have been conducted to leverage emerging IT technologies such as cloud computing, software defined networking (SDN), and network function virtualization (NFV), to accelerate this development. With the recent revolution in networking technology, these new communication mechanisms can open up more degrees of freedom in programmability and virtualization for tomorrow's grid.

5. **Cyber-Physical Security**

Cyber vulnerabilities of the power grid exist in numerous facilities, including Supervisory Control and Data Acquisition (SCADA) systems that power grids rely upon for monitoring and control of the physical grid. The increasing penetration of wind turbines, solar arrays, and energy storage devices creates potential new targets for cyber intrusions. About 65 million smart meters and other sensing devices, with capabilities for data collection for billing, outage reporting, and customers' choice, have been installed at the customer locations. These low-cost devices are targets for attackers, who may compromise large numbers of meters and use those meters to mount attacks. Such attacks resemble the October 2016 attack on the eastern U.S., in which smart devices such as wireless cameras were compromised and used to disrupt Internet access to millions of users. Ensuring resilience of the power grid to cyber threats will require a sustained and coordinated research effort towards understanding the vulnerabilities of cyber-attacks on the grid, providing defense mechanisms that can ameliorate the impact of attacks and ensure continued operation and efficient recovery from attacks when and where they occur. The cyber vulnerabilities of grid components must be characterized and mitigated. Privacy of sensitive power system data must also be preserved by efficient data protection and encryption techniques.

**Actions and Recommendations**

Based on the preceding discussion, we offer the following recommended actions:

1. Create a multi-university based national consortium on digital grid to jointly study the critical cyber-physical challenges based on a competitive solicitation.

2. Create a DOE led investment program to US industry and electric power companies to manufacture and upgrade the edge of the grid into a Digital Grid, such as the Smart Transformer manufacturing and deployment.

3. Enhance funding for programs such as the NSF's Cyber Physical System (CPS) program to focus on research needs associated with the Digital Grid.

4. Develop realistic and comprehensive testbeds for evaluations of control and cyber- security that would be impractical to conduct on the power grid itself.

5. Promote education and workforce development initiatives to train existing power engineers and grid operators about cyber-physical systems, data analytics, Internet-of-Things, cyber-security as well as educate the next generation of researchers and practitioners in the design and operation of secure and resilient power systems.

6. Form a task force by federal regulatory agencies to study the policy changes needed for creating a transactive energy market supported by the Digital Grid infrastructure.

**Acknowledgements**

The authors would like to acknowledge the following individuals for their contribution to this paper.

- Sean Meyn, University of Florida
- Anuradha Annaswamy, MIT
- Kevin Tomsovic, University of Tennessee
- Joe H. Chow, Rensselaer Polytechnic Institute
- Chen-Ching Liu, Washington State University
- Radha Poovendran, University of Washington, Seattle

*This material is based upon work supported by the National Science Foundation under Grant No. (1136993). Any opinions, findings, and conclusions or recommendations expressed in this material are those of the authors and do not necessarily reflect the views of the National Science Foundation.*